# What About Feedback?

By Terrence Letiche and Michael Lissack

November 27, 2014

Terrence Letiche, **thletiche@gmail.com**, Institute for the Study of Coherence and Emergence

Michael Lissack, **michael.lissack@gmail.com**, Executive Director, Institute for the Study of Coherence and Emergence and President, American Society for Cybernetics

**What About Feedback?**
November 27, 2014

*Abstract*

*The role of immediate feedback in-group conversations has received scant attention in the recent literature. While studies from the early 1990's suggested that "added information" in the form of non-verbal cues would allow video conferencing to "augment" the audio-only conference in terms of effectiveness, stunningly little follow-on research has been done reflective of the current state of computer mediated communication, video conferencing, "live walls", etc. This article contrasts three studies of immediate feedback in in-person settings as the basis for suggesting a new research program – research to look at potential effects of augmenting video-conferencing with an immediate feedback channel.*

1.  Introduction

Feedback plays a critical role in learning and in the overall management of systems. Feedback tends to be delivered after the fact – that is there is a time delay between when the events occur and are observed and when feedback is offered. When we engage with others in person by definition there are open immediate feedback channels – body language, eye movements, and , of course, direct speech. The greater the number of others with whom we are presently engaged and the more formal the setting, the smaller the attention we tend to give to such feedback. This tendency gets compounded when the modern technology of video conferencing is utilized. There the possibilities for immediate feedback get further reduced by limiting the open channels to audio and video. Efforts to "improve" the possibilities for feedback have been mostly concentrated on improving the quality of the audio and video transmissions. This article raises a series of questions regarding a "next step" – what if an immediate feedback channel was directly added into a video conferencing solution.

Throughout the 1950's to 1980's cybernetics, which many define as the study of feedback, was a leading academic discipline. But just as technology was enabling group meetings via audio and video, both cybernetics and the study of the role of feedback in group meetings faded into remission. What the studies and experiments into feedback systems by; Ashford & Northcraft (1992), Ashford & Tsui (1991), Morrison & Bies (1991) and Pritchard et al (1988) showed was that feedback is important in order to gauge problems and find solutions to them.

These late 20[th] century experiments and projects were goal orientated, meaning: let us investigate and use the implication of a feedback system within social-communicative situations to achieve a higher goal: increased efficiency of effectiveness (Pritchard et al., 1988). This research suffered from a need to measure key properties by implementing numerical values thus changing qualitative data into quantitative data and in effect superimposing a positivistic methodology on top of what was essentially contextual and constructivist data. (See. for example, two more recent papers on feedback systems: Strum et al., 2007 and Byun et al., 2011).

This article and its associated research agenda begins where those prior studies left off: with an assertion that the constructivist context of group meetings needs to be recognized and incorporated within the research design if the goal of "improving effectiveness" is to be addressed. The inherent social situations where feedback systems (such as those described herein) are designed to be used are, we would assert, best investigated through qualitative data achieved through narrative methodologies with quantitative data in a supporting role, rather than focusing on purely quantitative data achieved through positivistic methodologies. Meeting effectiveness is, we would assert, highly context dependent, and, as such, research methods which attempt to bracket that context dependence are unlikely to produce meaningful results.

In any form of interaction or conversation there are generally unspoken rules and protocols to which almost all participants abide. For instance, it is rude to interrupt someone while that person is speaking. Or it is rude to stare when approaching someone whom you have not met before, while it is, however, expected that you introduce yourself. A lot of these rules are conventions that enable us to converse and discuss things with others in an organized manner, making it possible for everyone to follow what is going on.

One approach to analysis, the Garfinkel tradition of social constructivism, focuses on these matters of social convention. According to Garfinkel (1984), the rules of social interaction are and need to be implicit; just like in riding a bicycle, the rules work just as long as we do not think about them. Thus the study of the social rules of interaction requires making the implicit, explicit. But because we need many implicit rules to interact, making these rules explicit makes the researched very uncomfortable. Garfinkel named making the implicit, explicit: 'breaching'. In conversation the participants constantly judge one another's participation, but only implicitly. It is just not done, minute-by-minute, to tell others in a conversation what you think of what they say. But for the conversation to occur, you have to attend to what the other says and respond to it; i.e. you have to constantly implicitly evaluate the speakers and what they say. In this experiment that necessary implicit process, is made explicit. (Heritage, 2001)

Some of the implicit rules or conventions are bothersome and can create problems; for instance, by confusing volume with participation allowing the loudest person to dominate, or even worse to bully by out-shouting others. Or by assuming that just because someone is quiet that they are uninvolved and have no opinion on the discussed topic or have nothing worth saying. This can be even truer if the forum is shifted to on-line from in-person. Saunders and Ahuja (2006) noted that "ongoing distributed teams are more difficult to manage and experience greater variance in well-being outcome levels" than do temporary teams. Johnson et al (2009) observed that team members who used computer-mediated communication more often experienced lower levels of positive affect while working with their teams and had lower levels of affective commitment to their teams. Together these observations highlight the import feedback can have – if it is meaningful, well timed, and collectively understood.

The three experiments highlighted below explore what happens in in-person settings when some general/social conventions (in regards to group discussion) are frustrated, circumvented, bypassed, thrown out, or just plainly demolished; whilst (thereby) other rules are strengthened, highlighted, emphasized and/or reimagined. Understanding the way people generally react to common situations and the underlying implicit rules makes it easier to anticipate behavior, allowing for individuals (in some part) to be steered or influenced.

The making of the implicit explicit is used in comedy --- for instance, there are clowns who mirror the actions of by-passers on a town square, creating a comic effect by playing with the unease that results from the clown's mime. A much less innocent use of such knowledge is when in marketing marketers try to influence decision making processes to decrease the barriers against spending money and to increase the chances of making sales (Barros, 1998). In these two examples, understanding of general/social conventions is manipulated to achieve a goal.

Each of the projects examined below purposely created an imbalance in the perception and/or perspective of the participants. Aspects of the implicit norms of group process/discussion were broken, whereby an alternative norm for the group process/discussion was made possible. Normally how much one speaks is not visible, tactile skills are irrelevant to business decision-making, and turn-taking in speech is determined hierarchically; thus by altering accepted forms of communication, the experimenters claimed to create an environment where productivity/effectiveness and creativity within the group, could (supposedly) thrive. One wishes to question this claim. The projects are:

1) *Reflect*; Bachour (2010), Bachour et al. (2008) and Kaplan (2012), which problematized group processes by adding a visual dimension of speech-controlled lights to interaction and communication.
2) *Simulation*; Mairesse (2014), which disrupted interaction and communication by separating listening and speaking from one another.
3) *TiC*; Letiche (2014), adds constant real-time participant controlled evaluative visual feedback to the discussion situation to invoke awareness of group dynamics. With

TiC, the goal of the research was to see if intense evaluative feedback will act as an inhibitor leading to stultified conversation and/or encourage participants to speak out (more) creatively.

2     The Experiments

The projects and experiments introduced above are explorative in nature. Each project confronts the existing forms of communication; there are no set goals other than to see what the outcome might be. Some of the projects might be considered to be more directed than others.

*2.1     Reflect*

*Reflect* is the product of the 2010 Doctoral dissertation of Khaled Bachour, with Pierre Dillenbourg and Frédéric Kaplan as supervisors. In essence, *Reflect* is a mirroring tool designed for use in a discussion/meeting setting. *Reflect* mirrors the vocal activity of participants in a group discussion. *Reflect* detects audible participation, the system then analyses the detected audio, and in turn the system produces visual feedback through the lighting up of LED's (located in the surface of the table around which the participants sit). More participation means that your color lights up more; less participation means your color lights up less. It was assumed that the participants would respond to the lights. Since the feedback produced directly corresponds to each individual's amount of audible participation, the system supposedly mirrors participation.

Visualization in the Reflect system is called "*Territorial*". This refers to how each participant's designated color does or does not spread across the surface of the table. More LED's light up as the amount of vocal participation by any particular individual increases. When that individual's participation decreases, the lit area shrinks.

*Reflect* creates a real-time feedback loop based on the level of audio detected by an automatic system. *Reflect* adds a visual dimension to people interacting and communicating. It makes the participation of each person in the discussion in real-time, i.e. as discussion is taking place, visible. The implicit --- varying levels of participation --- is made explicit.

New technology can be used in social and communications research to disrupt implicit assumptions and to see what effect(s) this can have. But when the reliance on technology outweighs/replaces/ignores social psychological/ sociological knowledge, the experiments are threatened to be naïve and/or irrelevant. "*Reflect*" suffers from such technological overindulgence. The quantity of speaking is definitely not seen in social science as the sole or key criteria for the quality of group participation. For instance, Belbin's "monitor/evaluator" and "completer/finisher" (Fisher, Hunter & Macrosson, 1998) speak little but are crucial to group effectiveness.

One lesson from *Reflect* is that the analytical focal point needs to be at least partially placed at the participant level to acknowledge that human judgment is crucial. .

*2.2     Simulation*

*Simulation* is an experiment designed by Philippe Mairesse in 2001 and described in his Doctoral dissertation in 2014; it is an experiment in communication, the intention of which is about the interactive quality of communication and not about delivering any specific content.

*Simulation* is intended for a large group of +/-12 people, sitting at a table connected to a system of microphones and headphones and switch boxes, all intertwined and connected to each other through cables to a single control-box. Because of this, it looks like a creature with a main body and 12 tentacles. Therefore, it was dubbed "the octopus". The octopus controls all the connections set by the 12 switchboxes. Each headset with microphone is connected to a switchbox with one dial to change between the listening-channels and a LED to indicate if the system is on.

The microphone is set to one channel and the headphones are connected to the switch in the switch box, which allows the participant to change between channels. The system allows one to speak to as many listeners as are tuned in to your channel, although you have no idea how many there are. There is a way to have a one-to-one conversation, but this only happens on the occasion that "participant one" is listening to the channel of "participant two", while "participant two" is listening to the channel of "participant one", although an unknown number of participants could still be listening in.

In normal conversational situations, you look at someone and there is a whole system of conventions on how to start a conversation. *Simulation* stops these from functioning, thus creating a very strange situation. It forces the participants to relinquish a certain amount of conventional control, normally present in the general communicative situation. Due to the disruption in communication conventions, the usual hierarchical control within a group does not occur; i.e. wherein teacher speaks in front of class, meaning that the students have to listen, and if a student doesn't listen the teacher tells him/her to be silent and to pay attention. This is a normal situation where authority plays a major role. But in the sessions of *Simulation*, this is made totally impossible.

Mairesse's hypothesis was that the situation would stimulate creative thought by reducing authority. The hypothesis made by Mairesse appears to be a partly correct one. *Simulation* does circumvent many aspects of authority. Although the stimulation of creativity may be more dependent on the willingness of the participants rather than on the disruption in communication created by *Simulation* and thus the effects of reducing hierarchical structure are less than certain.

For instance, in one session that took place at The Centre Pompidou in conjuncture with the Institut Télécom, Paris, the experimentation led to a verbal confrontation between one of the participants and one of the co-organizers of the colloquium during the debriefing portion of the experiment at the Making Sense Colloquium (2010). The confrontation was made possible due to the way *Simulation* demolished authority, allowing a participant to voice issues, which without the use of *Simulation*, would have gone unspoken and/or unnoticed to most present.

Simulation provided a break from authority, allowing for the occurrence of something unexpected. One lesson here was that if the "technological system" is to lead to creativity, the participants have to be willing to exploit the disruption of implicit speech hierarchy and rules, created by that very same system.

2.3 *TiC*

*TiC* was developed in 2014 by Terrence Letiche to build upon *Simulation* in the form of a communication group-experiment where the effect of a direct-feedback loop on group dynamics is investigated. The *TiC* experiment is process focused; the participants have been given the task of evaluating the quality of individual participation in the discussion, and not whether or not they agree with the evaluated participant's opinions. Within a discussion, people will have different opinions, and the articulation of these differences is what makes the discussion lively/interesting; and, in fact, possible. If everyone agrees, there is nothing to be said, and there won't be a reason to have discussions. Therefore it is not the goal of the experiment to measure (dis-)agreement among the participants; instead it is the character of the participation that is of interest and the attempt is to have that visualized.

The *TiC* intervention system consists of a discussion table with eight seating positions. For each position, from above by means of a projector, a uniquely colored ball shape is projected onto the surface of the discussion table visible by all participants. Every ball is directly in front/assigned to a seating position; each independent ball is adjustable in both size and brightness, these two variables are linked. Thus when a ball grows in size, the brightness increases and visa-versa. At each position there is a corresponding foot pedal, when the system is initiated it gives the pedals control over the adjustable variables of the balls. One pedal controls one ball resulting in a loop of linkages; A controls F, F controls D, D controls B, B controls H, H controls C, C controls E, E controls G and G controls A; the system does not allow control to be assigned to direct neighbors or opposite-facing participants, this is done to increase ambiguity among the participants and to avoid socially desirable behavior

(hypothesized as participants altering their behavior in such a way to conform to norms of conviviality, hoping to please others they believe are assessing them, in order to get desired results). While the evaluator is presumed to be making the light brighter/ bigger when he/she perceives the quality to be better -- and dimmer/ smaller when quality declines -- it should be noted that the participants are not given any notion of either scale or pre-assigned meanings with regard to what either size or brightness "mean." The ambiguity thus created was deliberate (though it itself creates limitations regarding inferences which may be drawn from the experimental sessions themselves) – because the research question was limited to "does feedback make a difference." Further experimentation with increased specificity of "meaning" with regard to the feedback mechanism would be necessary if the research question were to be narrowed so as to look at impact on effectiveness.

Three *TiC* sessions (discussions without intervention followed immediately by a continuation of that same discussion with the intervention present) with different groups were examined. Each session was coded by several coders with respect to participants' intensity of involvement, intensity of emotion, body language, and the leadership (high)/passivity (low) role being asserted or assumed. Each session began without the intervention mechanisms in place and then the interventions were added and the discussion session continued. Amongst sessions 1, 2 and 3, large differences were observed in the content analysis of the video record. The result of the intervention portion of the experiment each time was to push all the coded values more to the extreme, but in session 1 that was to higher levels and in 2 and 3 to lower ones. The *TiC* experiment produced a common group reaction towards more involvement and activity in group 1, while the very high cognitive level of the conversation before the intervention in session 2 dropped dramatically when the feedback started. The intervention in session 3 led to divisions and splintering in the group. The intervention by adding participation feedback loop in real-time had, in session 1, a collective effect towards more involvement, in session 2 it lead to more chaos (lessening of listening, poorer interaction and even weaker leadership), and in session 3 it had a divisive effect with a few participants becoming more dominant and even more participants becoming more passive.

Reflecting on the *TiC* data from a social studies perspective: there exists an implicit social rule or convention that in conversation you don't give (too much) critical feedback. Group maintenance needs often include displays of acceptance, in order to avoid conflict and group disintegration (Bales, 1958). Within *TiC*, this rule is marginalized or even abolished; the participants are intentionally placed in a situation where feedback is demanded of them, therefore potentially increasing tensions within the group. The fact that group maintenance conventions exist and that *TiC* succeeds in disrupting them becomes apparent when the difference in behavior/attitude by the participants is noticeable once the feedback intervention is introduced.

*2.4     The common denominator*

With each of the projects described, participant reactions and actions to altered social conventions were studied and commonalities observed. In each, it is clear that group dynamics change once the intervention or breaching was introduced.  As constructivist experiments, they each highlight the importance of participant processing of the intervention as part of the continuing interaction. For instance, it is not the lights, as such, of *Reflect* that tell people that they are keeping (very) quiet or that they are supposedly speaking too much; it is by observing the lights that the participants become acutely aware of their behavior, and their own position within their social surroundings. It is that awareness that counts.  (The realist would argue that the lights do communicate the extent of the participant's involvement and that the awareness is part of the reaction which evokes the possibility of change.  But, that dichotomy in perspective is not the subject of this article.)  In *Simulation*, it is not the disruption in the communication that (perhaps) enables authority to be broken and creativity to flourish, it is the participants becoming aware that authority is ever-present, and by having this awareness that they can be liberated, to let creativity flourish. Within TiC it is not the visual feedback provided in the form of changing size and brightness of the balls of light that enables the disruption within the communication among the participants, it is the participant's awareness gained by acknowledging/acting on the feedback about their position and role within the group dynamics that created the disruption, allowing for possible deviations within the form of communication taking place.

None of these experiments sheds any definitive light on the potential role of an immediate feedback channel in small group meetings.  As is true with regard to the literature as a whole, the limited studies conducted to date, while inconclusive with respect to ascertaining effectiveness of immediate feedback on groups, do show immediate feedback can have a dramatic effective on a meeting's efficacy. Since meeting efficacy is the usual goal of a video conference or a computer mediated meeting, what this observation suggests is that there may be a fruitful line of research into the efficacy effects of providing immediate feedback in a video conference or computer mediated small group meeting.  The technological options for running similar tests as these three "in-person" interventions in a video-conferencing setting are much greater. In the worst case scenario such research will determine immediate feedback in such a setting has no effect or even a detrimental effect on group dynamics. To the converse, a positive determination could open the door to a dramatic paradigm shift in the way video conferencing is facilitated.

3.      Questions Raised

While, with respect to the provision of immediate feedback to meeting participants, the existing literature base is scant, there are a set of questions and concerns therein which all three experiments need to address. These can be characterized as "participant cognitive load", "participant roles", and "participant preparedness/resistance"

### 3.1   Problems in concentration and overcrowding

Direct immediate feedback seems to be potentially problematic. In effect, asking the participants to discuss, evaluate and to reflect on themselves might overload them. It can be argued, for example, that if all the participants evaluate another participant within the group, while at the same time actively participating in the discussion, they would in effect be asked to split their attention between participation and evaluating, making it impossible for them to properly focus on either. This would either cause the quality of the participation to deteriorate, or endanger the quality of the evaluating process, for most people are incapable of splitting their attention in such a manner. The lesson here is that any attempts to obtain immediate feedback from fellow participants need to be compared to that from external observer(s).

### 3.2   Participants as recyclables

Most, if not all of the experiments in line with the "*breaching experiments*" by Garfinkel, suffer from a short lifespan. The reliability of the data can only be ensured with new "virgin" participants, who have no prior knowledge, understanding or expectations concerning the experimental system. As participants gain understanding of the experimental system's intent via participation, the participants become able to anticipate the events and/or deviations occurring due to the system, thus gaining knowledge on how to influence the results. The factor of unexpectedness is lost and the participants normalize what were at first unanticipated inputs. This causes a loss in validity of the data. The 'breaching experiments' examine normal assumptions by studying how participants react when these assumptions are interrupted. When the element of surprise disappears, the experiment's ability to see participants' unguarded reactions disappears, making the data from additional sessions meaningless. Consequently, the data is only valid as long as the participants are 'naïve' in their responses.

The lesson here is that the "positivist" assumptions of statistical validity based on independent uncorrelated events make demands on the experiment, which may in turn "drown out" the relevant effects. It may be better to adopt a "constructivist" framework where the participants are known to one another, are used to meeting in a group session, and are engaged in a task or discussion which both is on-going over multiple sessions and to which they have a personal commitment. In such an experimental design, the intervention variable will be the feedback itself and validity can only inferred from repeated applications of the intervention across multiple groups.

*3.3    No-one is spontaneous these days, self-control all around*

All of the breaching experiments described consist of two parts: there is the experimental session and then, afterwards, there is the debrief/review/evaluation. Due to the linear nature of the setup of the experiments, the feedback provided by the participants in the review sessions is indirect. Meaning, it is often focused on a few key moments; which quickly become a generalization, or supposedly exemplary, of the experience. McGuire and Botvinick (2010) note that we all have

> "… a mechanism for evaluating the levels of cognitive demand associated with specific task contexts … People apply cognitive control in order to preserve performance when demand rises. The detection mechanism registers the negative valence of situations in which unmet demand occurs. It provides a basis for long-term knowledge, causing people to avoid returning to inefficient situations if all else is equal."

The split between interaction and realization within each of the described experiments gives the participants time to reflect on the situation; allowing them to make a whole/complete/sensible "story" out of their experience, whilst leaving out subtle/out of place/ missed occurrences. Due to this post factum factor, the feedback has the risk of losing its spontaneity/sincerity and of ignoring 'weak signals' by favoring the 'big picture'.

The more recent works echo the limitations described above. The Strum et al. (2007) experiment was an automated system that functions on the positivistic premise that for a "good" meeting to occur the duration of participation needs to be equal. The automated system measured the time durations of each of the separate variables and visually represents (mirrors) the outcome to the participants on the table's surface through projection as feedback – but in so doing asserted that equal participation time was a desired goal. Byun et al. (2011) demonstrated that thorough usability testing/prototyping seems to be crucial to ensure that the system will not be overly complicated, uninviting, unclear and or confusing in its use. Byun and Strum together illustrated that a feedback system which merely mirrors through quantitative data feedback with minimal qualitative analysis can lead to an incorrect image of the social dynamics within group situations. It is important to have a system that is as clear, simple and unobtrusive as possible and which provides easily understood and commonly interpreted feedback.

The overall lesson here is one, which builds upon the prior observations above: the observation of immediate feedback needs to be made the focus of the next set of experiments themselves. Regardless of the degree of self-control exhibited by participants, the cumulative observations immediate feedbacks may generate can produce sufficient data so as to be inferable by an external observer. Or then again, it may not. But we need further experimentation to address such a hypothesis.

4       Discussion and conclusion

The experiments described, while all Garfinkelian breaching in nature, suggest a different path forward for further research. To our minds, that path lies not in the realm of in-person face-to-face interactions but rather in the realm of the virtual – the video conference. The protocol would be to adapt *TiC* and *Simulation* to video conferencing and embed the feedback technology into the "regular" interface used by the video conference participants.

The goal would be to see if immediate feedback can help render a video conference as effective as an in-person meeting. The challenges are many. Beck, et al, (2005) noted that "group performance [seems] better in face-to-face than in computer-mediated groups." Adams et al (2005) found that "familiarity plays a larger role in computer mediated groups than in face to face groups." These studies built upon the Burke et al (1999) observation that "groups using richer media felt their medium was more effective." Thus, the question: can the provision of an immediate feedback channel enrich the video conferencing environment enough to create effectiveness as an affordance?

Both *TiC* and *Simulation* can be regarded as falling between an interactive art installation and a social psychological breaching experiment. The suggested protocol would allow their extension into the domains of Human Computer Interaction and Computer Mediated Communication. While both HCI and CMC tend to focus on psychological reaction to technology, the research protocol stemming from our observation of these experiments would instead use media technology to study human sociability. Waller et al (2013) noted that "the ability of group members to shape group behaviors can greatly influence collective outcomes; however, the skills associated with correctly recognizing behaviors in situ and responding appropriately to them on a real-time basis are typically not emphasized in group dynamics education." One objective of the research would be to attempt to alter that very dynamic.

The protocol itself is rather simple and direct: one or more meeting participants will be provided with feedback in the form of a colored dot (or light bulb etc.) to be displayed as part of the participant's "self" panel in the video conferencing display. The observer supplying the feedback will be given a set of slider controls which control color size and brightness of the dot shown to the participant. The feedback itself will be focused on the manner of that participant's participation: quantity of discussion relative to others and to the participant's expected role, emotional intensity being displayed, "fidgetiness" and other forms of body language, and perhaps the use of inappropriate language or outbursts. Unlike the Strum experiment, accumulated time is not the measure, but instead it will be time spent relative to the self-identified nature of the role within the meeting/discussion/conference by the participants involved together with feedback produced from within the group. (Thus, for example, if the ROYGBIV color scale were to be used, red might indicate "please stop" and purple, "participate more" with the size and intensity of the colored dot conveying the intended intensity of the feedback.) What will not be examined is the actual content of the participation. The protocol will call for varying numbers of meeting participants to be given such feedback and will examine the differences if any between feedback derived from an external observer (not a participant in the meeting) and feedback derived from another meeting participant.

On-going meetings and a variety of feedback channel displays and interventions will be attempted and narratives regarding their success/failure recorded. Such interventions may become "expected" by the participants and the effects of those expectations need to be incorporated into the constructed narrative (see Bremond, 1973, and Cavazza et al, 2010). At each stage the research must keep in mind an important warning:

> Although people respond socially to computer products that convey social cues, to be effective in persuasion, designers must understand the appropriate use of those cues. …when you turn up the volume on the "social" element of a persuasive technology product, you increase your bet: you either win bigger or lose bigger, and the outcome often depends on the user. If you succeed, you make a more powerful positive impact. If you fail, you make users irritated or angry. (Fogg, 2003)

Accepting the context of the group meetings as non-random and non-independent will be a key component of the research protocol. It makes little sense , at least to us, to assume that any improvements in meeting effectiveness would occur when the meetings examined involved randomly gathered groups discussing random topics. While the experiments of the 80's and 90's attempted to examine the impacts of feedback on random groups of independent people gathered solely for the purpose of the experiment itself, we believe that it is critical to provide a context where it is the feedback and its form which is the independent variable and where group norms, cohesion, topic of meeting, and purpose are all accepted as "norms". This means the groups to be examined will be pre-existing and used to regular video meetings around a pre-established agenda. This will eliminate the Byun error of participants who reported confusion over the meaning of the feedback. It also means that some measure of the "effectiveness" of the intervention(s) will be observable by the participants themselves. (And, if there is such self-reporting by participants, is likely to then create a positive feedback loop regarding the perceived "value" to the group of a particular intervention.) However, the use of pre-established groups will, by definition, eliminate the possibility of positivistic statistical inference (until the research set acquires the property of large numbers), and as such, the conclusions to be reached will be more of a narrative ethnography of the changes in the groups and their meetings. That narrative, in turn, may then suggest avenues for further study.

From the perspective of the software suppliers who provide video conferencing solutions, in an ideal world, feedback mechanisms generated by external observation will be shown to have positive results on the perceived efficacy of the meeting. If so then the potentiality for computer assisted communication in this realm can augment the current roles for computer-mediated communication. Of course, the opposite effect may turn out to occur which would suggest a reason for such firms to avoid implementing such assistance regimes. It may also be the case that the positive efficacy effects are generated but are then mitigated by some form of Mori Uncanny Valley (Mori, 1970) effect where the meeting participants feel "creeped out" by the "big brother is watching/listening" aspects of the assistance regime so deployed. Or perhaps, the attentional demands created by the presence of the immediate feedback will have an opposite effect on efficacy. And, as in any study, which is testing hypotheses de novo, it may be that in the limited realm of small groups meeting via video conferencing that immediate feedback is not meaningful at all. However, given the increasing ubiquity of the video conference small group meeting it clearly is worthwhile to attempt to discover what relationships may hold.

Conclusion and Next Steps

Further research re the use immediate feedback channel(s) needs to head in the direction of activity theory centered HCI. As Bertelsen and Bedker (2002) describe it, the goal should be a: "focus on actual use and the complexity of multiuser activity. In particular, the notion of the artifact as mediator of human activity is essential." From this perspective, the gaps in the literature regarding immediate feedback in the context of small group meetings seem to be conspicuous. Placing those gaps in the context of the constraints imposed by a demand for positivistic statistical inquiry helps to explain why the research stopped. But, there is no reason for those positivistic statistical inference demands to be governing. Much of the recent literature and research on the role of feedback in learning has taken a constructivist narrative ethnographic approach and there is no reason that such an approach should not be attempted with the realm of video conferencing and small group meetings. The research protocol suggested herein appears to be a logical next step in determining if the obstacles to effectiveness for video conferencing highlighted by Adams, Beck, Burke, Waller can be overcome.